\documentstyle[prl,aps,multicol]{revtex}

\begin{document}

\title{Quantum Gambling}
\author{Lior Goldenberg, Lev Vaidman and Stephen Wiesner}
\address{School of Physics and Astronomy \\
Raymond and Beverly Sackler Faculty of Exact Sciences \\
Tel-Aviv University, Tel-Aviv 69978, Israel.}
\date{August 2, 1998}
\maketitle

\begin{abstract} 
  
We present a two-party protocol for quantum gambling, a new task closely related to coin 
tossing. The protocol allows two remote parties to play a gambling game, such that in a certain 
limit it becomes a fair game. No unconditionally secure classical method is known to accomplish 
this task.

\end{abstract}

\begin{multicols}{2}

  
Quantum cryptography is a field which combines quantum theory with information theory. The goal 
of this field is to use the laws of physics to provide secure information exchange, in contrast 
to classical methods based on (unproven) complexity assumptions. In particular, quantum key 
distribution protocols \cite{QKD} became especially important due to technological advances 
which allow their implementation in the laboratory. However, the last important theoretical 
result in the field was of a negative character: Mayers \cite{Mayers} and Lo and Chau \cite{LC1} 
showed that quantum bit commitment is not secure. Their work also raised serious doubts on the 
possibility of obtaining any secure two-party protocol, such as oblivious transfer and coin 
tossing \cite{Lo}. In this Letter we present a secure two-party quantum cryptographic task -- 
quantum gambling, which has no classical counterpart.
  
Coin tossing is defined as a method of generating a random bit over a communication channel 
between two distant parties. The parties, traditionally named Alice and Bob, do not trust each 
other, or a third party. They create the random bit by exchanging quantum and classical 
information. At the end of the protocol the generated bit is known to both of them. If a party 
cheats, i.e. changes the occurrence probability of an outcome, the other party should be able 
to detect the cheating. We would consider a coin tossing protocol to be secure if it defines a 
parameter such that when it goes to infinity the probability to detect any finite change of 
probabilities goes to $1$. Using a secure protocol the parties can make certain decisions 
depending on the value of the random bit, without being afraid that the opponent may have some 
advantage. For instance, Alice and Bob can play a game in which Alice wins if the outcome is `0' 
and Bob wins if it is `1'. Note that if bit commitment were secure, it could be used to 
implement coin tossing trivially: Alice commits a bit $a$ to Bob; Bob tells Alice the value of 
a bit $b$; the random bit is the parity bit of $a$ and $b$.
  
It is not known today if a secure quantum coin tossing protocol can be found \cite{RCT}. It is 
only known that {\it ideal} coin tossing, i.e. in which no party can change the expected 
distribution of the outcomes, is impossible \cite{LC2}. Based on our efforts in this direction, 
we are skeptical about the possibility to have secure (non-ideal) coin tossing. Nevertheless, 
we were able to construct a protocol which gives a solution to a closely related task. 
``Quantum gambling'' is very close to playing in a casino located in a remote site, such as 
gambling over the Internet. As in a real casino, for instance when playing Roulette, the 
player's possible choices give him some probability to win twice the amount of his bet, or a 
smaller probability to win a bigger sum. However, in our protocol the player has only a partial 
control over these choices. In spite of its limitations our protocol provides a quantum 
solution to a useful task, which cannot be performed securely today in the classical framework.  
Assuming ideal apparata and communication channels, the protocol is unconditionally secure, 
depending solely on the laws of physics.
  
Let us start by defining exactly the gambling task considered here. The casino (Alice) and the 
player (Bob) are physically separated, communicating via quantum and classical channels. The 
bet of Bob in a single game is taken for simplicity to be $1$ coin. At the end of a game the 
player wins $1$ or $R$ coins, or loses 1 coin (his bet), depending on the result of the game. 
We have found a protocol which implement this game while respecting two requirements: First, 
the player can ensure that, irrespective of what the casino does, his expected gain is not less 
than $\delta$ coins, where $\delta$ is a negative function of $R$ which goes to zero when $R$ 
goes to infinity. The exact form of $\delta (R)$ will be specified below. Second, the casino 
can ensure that, irrespective of what the player does, its expected gain is not less than $0$ 
coins.

In order to define the protocol rigorously, we will first present the rules of the game, then 
the strategies of the players which ensure the outcomes quoted above and finally we will prove 
the security of the method.

{\bf The Rules of the Game:} 
Alice has two boxes, $A$ and $B$, which can store a particle. The quantum states of the 
particle in the boxes are denoted by $|a\rangle$ and $|b\rangle$, respectively. Alice prepares 
the particle in some state and sends box $B$ to Bob.

Bob wins in one of the two cases:
\begin{enumerate}
\item
If he finds the particle in box $B$, then Alice pays him $1$ coin (after checking that box $A$ 
is empty).
\item
If he asks Alice to send him box $A$ for verification and he finds that she initially prepared 
a state {\it different} from 
\begin{equation}
|\psi_{0}\rangle = \frac{1}{\sqrt{2}} \: (|a\rangle + |b\rangle) ,
\label{Psi_0}
\end{equation}
then Alice pays him $R$ coins.
\end{enumerate}
In any other case Alice wins, and Bob pays her $1$ coin.

The players' strategies which ensure (independently) an expectation value of Alice's gain  
$G_A \geq 0$ (irrespective of Bob's actions) and an expectation value of Bob's gain 
$G_B \geq \delta$ (irrespective of Alice's actions) are as follows:

\noindent
{\bf Alice's  Strategy:} 
Alice prepares the equally distributed state $|\psi_{0}\rangle$ (given in eq.(\ref{Psi_0})).

\noindent
{\bf Bob's  Strategy:} 
After receiving box $B$, Bob splits the particle in two parts; specifically, he performs the 
following unitary operation:
\begin{equation}
|b\rangle \rightarrow \sqrt{1 - \eta} \; |b\rangle + \sqrt{\eta} \; |b'\rangle,
\label{Bob-split}
\end{equation}
where $\langle b'|b\rangle = 0$. The particular splitting parameter $\eta$ he uses is 
$\eta = \tilde{\eta}(R)$ (to be specified below). After the splitting Bob measures the 
projection operator on the state $|b\rangle$, and then
\begin{enumerate}
\item[I.]  
If the measurement yields a positive result, i.e. he finds the particle, he announces Alice 
that he won.
\item[II.]  
If the measurement yields a negative result, he asks Alice for box $A$ and verifies the 
preparation.
\end{enumerate}
This completes the formal definition of our protocol.

In order to prove the security of the scheme, we will analyze the average gain of each party 
as a result of her/his specific strategy. It is straightforward to see that Alice's strategy 
ensures $G_A \geq 0$. If Alice prepares the state $|\psi_{0}\rangle$, Bob has no meaningful way 
of increasing his odds beyond $50\%$: if he decides to open box $B$ he has a probability of 
$0.5$ to win $1$ coin and a probability of $0.5$ to lose $1$ coin. He cannot cheat by claiming 
that he found the particle when he did not, since Alice learns the result by opening box $A$. 
If, instead, he decides to verify the preparation he will find the expected state, so he will 
lose 1 coin. Therefore $G_B \leq 0$, and since this is a zero-sum game, Alice's gain is 
$G_A \geq 0$, whatever Bob does.

Now we will prove that Bob, using the splitting parameter $\eta = \tilde{\eta}$, can ensure 
$G_B \geq \delta$. The values of $\tilde{\eta}$ and $\delta$ are determined by the calculation 
of Bob's expected gain, $G_B$. Bob tries to maximize $G_B$ under the assumption that Alice uses 
the worse strategy for him, namely the one which minimizes $G_B$ for Bob's particular strategy. 
Therefore, we will first minimize the function $G_B$ for any $\eta$, and then we will find the 
maximum of the obtained function, with that computing $\delta$. We will also compute the value 
of $\eta$ at the peak, $\tilde{\eta}$, which will be the chosen splitting parameter of Bob.

Let us first write down the expression for $G_B$. Bob gets $1$ coin if he detects the state 
$|b\rangle$; denote the probability for this event to occur by $P_b$. He gets $R$ coins if he 
detects a different preparation than $|\psi_{0}\rangle$ (after failing to find the state 
$|b\rangle$, an event with a related probability of $1 - P_b$); denote the probability to 
detect a different preparation by $P_D$. He loses 1 coin if he does not detect a different 
preparation than $|\psi_{0}\rangle$ (after failing to find $|b\rangle$); the probability for 
this event is $(1 - P_D)$. Thus, the expectation value of Bob's gain is 
\begin{equation}
G_B = P_b + (1 - P_b) \; [P_D R - (1 - P_D)] . 
\label{G_B-def}
\end{equation}

For the calculations of $P_b$ and $P_D$ we will consider the most general state Alice can 
prepare. In this case the particle may be located not only in boxes $A$ and $B$, but also in 
other boxes $C_i$. The states $|a\rangle$, $|b\rangle$ and $|c_i\rangle$ are mutually 
orthogonal. She can also correlate the particle to an ancilla $|\Phi\rangle$, such that the 
most general preparation is 
\begin{equation}
|\Psi_0\rangle = \alpha |a\rangle |\Phi_a\rangle + \beta |b\rangle |\Phi_b\rangle + 
\sum_i \gamma_i |c_i\rangle |\Phi_{c_i}\rangle ,
\label{Psi_0'}
\end{equation}
where $|\Phi_a\rangle$, $|\Phi_b\rangle$, $|\Phi_{c_i}\rangle$ are the states of the ancilla
and $|\alpha|^2 + |\beta|^2 + \sum_i |\gamma_i|^2 = 1$. After Bob splits $|b\rangle$, as 
described by eq.(\ref{Bob-split}), the state changes to
\begin{eqnarray}
|\Psi_1\rangle & = \alpha |a\rangle |\Phi_a\rangle + \beta & \left( \sqrt{1 - \eta} \; 
|b\rangle + \sqrt{\eta} \; |b'\rangle \right) |\Phi_b\rangle \nonumber \\
& & + \sum_i \gamma_i |c_i\rangle |\Phi_{c_i}\rangle .
\label{Psi_1'}
\end{eqnarray}
The probability to find the state $|b\rangle$ (in step I. of Bob's strategy) is
\begin{equation}
P_b = ||\langle b|\Psi_1\rangle||^2 =  |\beta|^2 \: (1 - \eta).
\label{P_b}
\end{equation}
If Bob does not find $|b\rangle$, then the state reduces to
\begin{equation}
|\Psi_2\rangle = {\cal N}\left( \alpha |a\rangle |\Phi_a\rangle + \beta \sqrt{\eta} \; 
|b'\rangle |\Phi_b\rangle + \sum_i \gamma_i |c_i\rangle |\Phi_{c_i}\rangle \right) ,
\label{Psi_2'}
\end{equation}
where ${\cal N}$ is the normalization factor given by ${\cal N} = \left( 1 - (1 - \eta) \: 
|\beta|^2 \right)^{-1/2}$. On the other hand, if Alice prepares the state $|\psi_0\rangle$ 
instead of $|\Psi_0\rangle$, then at this stage the particle is in the state
\begin{equation}
|\psi_2\rangle = \sqrt{\frac{1}{1 + \eta}} \; |a\rangle + \sqrt{\frac{\eta}{1 + \eta}} \; 
|b'\rangle .
\label{Psi_2}
\end{equation}
Thus, the best verification measurement of Bob is to make a projection measurement on this 
state. If the outcome is negative, Bob knows with certainty that Alice did not prepared the 
state $|\psi_0\rangle$. The probability of detecting such a different preparation is given by
\begin{eqnarray}
P_D  & = & 1 -  ||\langle\psi_2|\Psi_2\rangle||^2 \nonumber \\
& = & 1 - {\cal N}^2 \; \left|\left| \frac{\alpha}{\sqrt{1 + \eta}} |\Phi_a\rangle +
\frac{\beta \eta}{\sqrt{1 + \eta}} |\Phi_b\rangle \right|\right|^2 .
\label{P_D}
\end{eqnarray}

Since Alice wants to minimize $G_B$, she tries to minimize both $P_b$ and $P_D$. From 
eq.(\ref{P_D}) we see that in order to minimize $P_D$, the states of the ancilla 
$|\Phi_a\rangle$ and $|\Phi_b\rangle$ have to be identical (up to some arbitrary phase), i.e.
$|\langle\Phi_a|\Phi_b\rangle| = 1$. That is, Alice gets no advantage using an ancilla, so it 
can be eliminated. Then, in order to maximize ${\cal N} |\alpha + \beta \eta|$, Alice should 
set all $\gamma_i$ to zero, as it is clear from the normalization constraint $|\alpha|^2 +
|\beta|^2 = 1 - \sum_i |\gamma_i|^2$. This operation has no conflict with the minimization of 
$P_b$, since eq. (\ref{P_b}) contains only $|\beta|$. Also, the maximization is possible if the 
coefficients $\alpha$ and $\beta$, if seen as vectors in the complex space, point in the same 
direction. Therefore, Alice gains nothing by taking $\alpha$ and $\beta$ to be complex numbers; 
it is sufficient to use real positive coefficients. Taking all these considerations into 
account, the state prepared by Alice can be simplified to
\begin{equation}
|\psi_0'\rangle = \sqrt{\frac{1}{2} + \epsilon} \; |a\rangle + \sqrt{\frac{1}{2} - \epsilon} 
\; |b\rangle .
\label{psi_0'}
\end{equation}

Now, the state after Bob splits $|b\rangle$ reads
\begin{equation}
|\psi_1'\rangle = \sqrt{\frac{1}{2} + \epsilon} \; |a\rangle + \sqrt{\frac{1}{2} 
- \epsilon} \; \left( \sqrt{1 - \eta} \; |b\rangle + \sqrt{\eta} \; |b'\rangle \right) ,
\label{Psi_1''}
\end{equation}
and so the probability to find $|b\rangle$ becomes
\begin{equation}
P_b = ||\langle b|\psi_1'\rangle||^2 =
\left( \frac{1}{2} - \epsilon \right) (1 - \eta) .
\label{P_b'}
\end{equation}
When Bob does not find the state $|b\rangle$, $|\psi_1'\rangle$ reduces to
\begin{equation}
|\psi_2'\rangle = \frac{\sqrt{1 + 2 \epsilon} \: |a\rangle + \sqrt{\eta \: (1 - 2 \epsilon)}
 \: |b'\rangle }{\sqrt{1 + 2 \epsilon + \eta \; (1 - 2 \epsilon)}} ,
\label{Psi_2''}
\end{equation}
which in turn leads to
\begin{equation}
P_D  = 1 -  ||\langle\psi_2|\psi_2'\rangle||^2 =
\frac{2 \eta \left( 1 - \sqrt{1 - 4 \epsilon^2} \right)}
{(1 + \eta)^2 + 2 \epsilon  \; (1 -  \eta^2 )} .
\label{P_D'}
\end{equation}
Substituting eq.(\ref{P_b'}) and eq.(\ref{P_D'}) in eq.(\ref{G_B-def}), we find $G_B$ in terms 
of the splitting parameter $\eta$, the preparation parameter $\epsilon$ and $R$:
\begin{eqnarray}
G_B = - \frac{1}{1 + \eta} & & \left[
2 \epsilon \, (1 - \eta^2) + \eta \, (\eta + \sqrt{1 - 4 \epsilon^2}) \right. \nonumber \\
& & \;\; - \left. \eta \, (1 - \sqrt{1 - 4 \epsilon^2}) \; R \right] .
\label{G_B}
\end{eqnarray}

In order to calculate the minimal gain of Bob, $\delta$, irrespective of the particular 
strategy of Alice, we will first minimize $G_B$ for $\epsilon$ and then maximize the result for 
$\eta$:
\begin{equation}
\delta(R) = Max_{\, \eta} \, [Min_{\, \epsilon} \, G_B(R,\eta,\epsilon)] .
\label{delta-def}
\end{equation}
The calculations yield
\begin{eqnarray}
\delta = - & & \frac{1}{1 + \sqrt{R + 2 - \sqrt{(R + 2)^2 - 1}}} \nonumber \\
& & \times \left\{ 2 + \left[ R - \sqrt{(R + 2)^2 -1} \right] 
  \phantom{\sqrt{\sqrt{(R)^2}}} \right. \nonumber \\
& & \times \left. \left[ 1 - \sqrt{R + 2 - \sqrt{(R + 2)^2 - 1}} \right] \right\} ,
\label{delta}
\end{eqnarray}
obtained for Bob's  splitting parameter 
\begin{equation}
\tilde{\eta} = \sqrt{R + 2 - \sqrt{(R + 2)^2 - 1}} .
\label{eta_B}
\end{equation}
In the range of $R \gg 1$, these results can be simplified to
\begin{eqnarray}
\delta \approx - \sqrt{\frac{2}{R}} , 
\label{delta-R>>1} \\
\tilde{\eta} \approx \sqrt{\frac{1}{2R}} .
\label{eta_B-R>>1}
\end{eqnarray}

We have shown that if Bob follows the proposed strategy with $\eta = \tilde{\eta}$, then his 
average gain is not less than $\delta$; this bound converges to $0$, i.e. to the limit of 
a fair game, for $R \rightarrow \infty$. This is true for any possible strategy of Alice, 
therefore, the security of the protocol is established.

To compare our scheme to a real gambling situation, let us consider the well-known Roulette 
game. A bet of $1$ coin on the red/black numbers, i.e. half of the $36$ numbers on the table, 
rewards the gambler with $1$ coin once in $18/38$ turns (on average, for a spinning wheel with 
38 slots); this gives an expected gain of about $-0.053$ coins. To assure the same gain in our 
scheme, $R = 700$ is required. Note that extremely large values of $R$ are practically 
meaningless, one reason being the limited total amount of money in use. Nevertheless, the bound 
on $\delta$ is not too restrictive when looking at the first prizes of some lottery games: a 
typical value of $R = 10^6$ gives a reasonably small $\delta$ of about $-0.0014$.

It is also interesting to consider the case of $R = 1$. This case corresponds to coin tossing, 
since it has only two outcomes: Bob's gain is either $-1$ coin (stands for bit `0') or $1$ coin 
(stands for bit `1'). The minimal average gain of Bob is about $-0.657$, which translates to an 
occurrence probability of bit `1' of at least $0.172$ (instead of $0.5$ ideally), whatever 
Alice does. This is certainly not a good coin tossing scheme, however, no classical or quantum 
method is known to assure (unconditionally) {\it any} bound for the occurrence probability of 
both outcomes.

Our analysis so far was restricted to a single instance of the game, but the protocol may be 
repeated several times. After $N$ games Bob's expected gain is $G_B \geq N \delta$ and Alice's 
expected gain is $G_A \geq 0$. Of course, Alice may choose now a complex strategy using 
ancillas and correlations between particles/ancillas from different runs. In this way she may 
change the probability distribution of her winnings, but she cannot reduce the minimal expected 
gain of Bob. Indeed, our proof considers the most general actions of Alice, so the average gain 
of Bob in each game is not less than $\delta$, and consequently, it is not less then $N \delta$ 
after $N$ games. A similar argument is valid for Bob's actions, so the average gain of Alice 
remains non-negative even after $N$ games. In gambling games, in addition to the average gain, 
it is important to analyze the standard deviation of the gain, $\Delta G$. Bob will normally 
accept to play a game with a negative gain only if $\Delta G_B \gg |G_B|$ (unless he has some 
specific target in mind). In a single application of our protocol, $\Delta G_B \geq 1$, so the 
condition is attained for big enough values of $R$ (see eq.(\ref{delta-R>>1})). However, 
increasing the number of games makes the gambling less attractive to Bob: if Alice follows the 
proposed strategy, $|G_B|$ grows as $N$ while $\Delta G_B$ grows only as $\sqrt N$. Therefore, 
Bob should accept to play $N$ times only if $N \ll 1 / \delta^2 \sim R$.

Another important point to consider is the possible ``cheating'' of the parties. Alice has no 
meaningful way to cheat, since she is allowed to prepare any quantum state and she sends no 
classical information to Bob. Any operation other than preparing $|\psi_0\rangle$, as adding 
ancillas or putting more/less than one particle in the boxes, just decreases her minimal gain. 
Bob, however, may try to cheat. He may claim that he detected a different preparation than 
$|\psi_0\rangle$, even when his verification measurement does not show that. If Alice prepares 
the initial state $|\psi_0'\rangle$ (with $\epsilon > 0$), she is vulnerable to this cheating 
attempt: she has no way to know if Bob is lying or not. For this reason Alice's proposed 
strategy is to prepare $|\psi_0\rangle$ every time, such that any cheating of Bob could be 
invariably detected. When both parties follow the proposed strategies, i.e. $\epsilon = 0$ and 
$\eta = \tilde{\eta}$, the game is more fair for Bob than assumed in the proof:
\begin{equation}
G_{B_{\, prot}} = - G_{A_{\, prot}} =  - \sqrt{R + 2 - \sqrt{(R + 2)^2 - 1}} .
\label{G_A_prot,G_B_prot}
\end{equation}
For $R \gg 1$ we get $G_{B_{\, prot}} \approx - 1 / \sqrt{2R}$, which is approximately half of 
the value of $\delta$ calculated in eq.(\ref{delta-R>>1}).

The discussion up to this point assumed an ideal experimental setup. In practice errors are 
unavoidable, of course, and our protocol is very sensitive to the errors caused by the devices 
used in its implementation (communication channels, detectors, etc). In the presence of errors, 
if the parties disagree about the result of a particular run it should be canceled. If such 
conflicts occur more than expected based on the experimental error rate, it means that (at 
least) one party is cheating, and the game should be stopped. The most sensitive part to errors 
is the verification measurement of Bob, i.e. the detection of the possible deviation of the 
initial state from $|\psi_0\rangle$. In the ideal case, using $\tilde{\eta}$ and the 
corresponding $\epsilon$ (the worst for honest Bob), the detection probability is very small: 
$P_D \approx \sqrt{2/R^3}$, for $R \gg 1$. Clearly, for a successful realization of the 
protocol, the error rate has to be lower than this number. Thus, in practice, the experimental 
error rate will constrain the maximal possible value of $R$ \cite{Bob-strategy-2}.

In conclusion, we have built a simple yet effective protocol for quantum gambling. We have 
proved that no party can increase her/his winnings beyond some limit, which converges to $0$ 
when $R$ goes to infinity, if the opponent follows the proposed strategy. An important aspect 
of our protocol is that it shows that secure two-party quantum cryptography is possible, in 
spite of the failure of quantum bit commitment. The possibility of having other so-called 
``post-cold-war applications'' remains an open question.

We would like to thank H.K. Lo and D. Mayers for enlightening discussions on bit commitment and 
coin tossing. This research was supported in part by Grants No. 614/95 and 471/98 of the Israel 
Science Foundation. Part of this work was done during the 1996 and 1997 Elsag-Bailey Foundation 
research meetings on quantum computation. 


\end{multicols}


\begin{thebibliography}{99}
\bibitem{RCT}  Note that if we limit ourselves to spatially extended secure sites located one 
near the other, then secure coin tossing can be realized classically, by simultaneous exchange
of information at the opposite sides of the sites. The security of this method relies on 
relativistic causality.
\bibitem{Bob-strategy-2} In the presence of errors an alternative strategy of Bob, without 
splitting state $|b\rangle$, is more efficient: Bob requests box $A$ for verification randomly 
(at a rate optimized for the value of $R$) and measures the projection on $|\psi_0\rangle$ (not 
on $|\psi_2\rangle$). Although for a given $R$ this strategy offers him a lower gain ($\delta 
\approx - 2 / \sqrt{R}$, for $R \gg 1$), it allows using a higher $R$ for a given fidelity of 
the setup, since now $P_D \approx 1 / R$.
\end{thebibliography}
\end{document}